\documentclass[preprint,showpacs,preprintnumbers,amsmath,amssymb]{revtex4}

\usepackage{epsf}
\usepackage{graphicx}% Include figure files
\usepackage{dcolumn}% Align table columns on decimal point
\usepackage{bm}% bold math
\newcommand{\°}{$^\circ$}

\begin{document}
\title{Drift of charged defects in local fields as aging mechanism in ferroelectrics}
\author{Yuri A. Genenko} \email{yugenen@tgm.tu-darmstadt.de} %
\affiliation{Institute of Materials Science, %
Darmstadt University of Technology, %
Petersenstr. 23, 64287 Darmstadt, Germany}%

\author{Doru C. Lupascu}%
\affiliation{Institute of Materials Science, %
Dresden University of Technology, %
Helmholtzstra\ss e 7, 01069 Dresden, Germany}%
%\homepage{http://www.tu-darmstadt.de/fb/ms/fg/na/}%
\date{\today}

\begin{abstract}

Point defect migration is considered as a mechanism for aging in
ferroelectrics. Numerical results are given for the coupled problems of point
defect migration and electrostatic energy relaxation in a 2D domain configuration.
The peak values of the clamping pressure at domain walls are in the range of $10^6$\,Pa,
which corresponds to
macroscopically observed coercive stresses in perovskite ferroelectrics. The
effect is compared to mechanisms involving orientational reordering of defect
dipoles in the bulk of domains. Domain clamping is significantly stronger in
the drift mechanism than in the orientational picture for the same material
parameters.
\end{abstract}

\maketitle

\section{\label{sec:intro}Introduction}

Ferroelectric materials underlie restrictions in technological applications
because of several degradation phenomena. One of these phenomena is aging which
is defined as the gradual change of material properties with time $t$ under
static external boundary conditions
\cite{drougard54domain,mcquarrie55aging,takahashi70space,%
thomann72stabilization,carl78electrical,arlt88internal,zhang05insitu,zhang06aging}.
Experimentally, the dielectric constant decreases and hysteresis loops alter
their shape and amplitude in remanent polarization and coercive field. The
dielectric constant decreases as the logarithm of time in an intermediate time
range saturating for long times \cite{plessner56aging}. The change of material
properties is caused by a decreasing domain wall mobility stabilizing in an
aged domain structure \cite{ikegami67mechanism}. In order to describe the
experimentally observed shifted or deformed ferroelectric hysteresis loops
after aging \cite{lamb86_02} the internal field ${\bf E}^i$ has been defined as
a quantity describing the strength of domain stabilization
\cite{arlt88internal}. Several mechanisms have been considered to partake in
the stabilization process, space charge formation
\cite{takahashi70space,thomann72stabilization}, domain splitting
\cite{ikegami67mechanism}, ionic drift
\cite{hage80_02,lamb86_02,scott87activation}, or the reorientation of defect
dipoles \cite{robels93domain}. Except for domain splitting, all mechanisms
directly involve some reordering of point defects. Within a microstructure,
three relevant locations can be identified for charge carrier rearrangement:
the domain bulk, grain boundaries, or domain walls. For the bulk effect, defect
dipoles reorient with respect to the direction of the spontaneous polarization
under an electrical field or strain. For the grain boundary effect, charged
point defects move under electrical fields originating from polarization
discontinuities at the grain boundaries or the outer perimeter of the sample in
order to compensate the fields. The same process can occur at charged domain
walls and then becomes a domain wall effect
\cite{chynoweth60pyroelectricity,shur88spatial}. 
 Local space charge is another electric driving force for
ionic currents observed in LiNbO$_3$ type crystals
\cite{gopalan96observation,imlau03holographic} as well as perovskites
\cite{alemany84ageing,korneev01thermal,gakh01space}.

Elastic fields can provide a second driving force for defects inside domains
but will not be treated here. For not too rigid non-charged domain walls, the
localization of free charge carriers at a domain wall is a further possible
effect entailing the wrinkling of the initially planar walls
\cite{mueller02dielectric,mueller03aging,mueller04aging,paruch05domain}. Even
though this is a physically interesting mechanism, it will not be taken into
consideration here.

Oxygen vacancies are a well known and frequent defect in the perovskite
structure. They have been considered to play an important role in aging of
ferroelectric materials due to their low, but finite mobility. In the
orientational picture, oxygen vacancies, when adjoint to an acceptor center,
form electric and elastic defect dipoles in the bulk of a ferroelectric domain
\cite{arlt88internal}. The defect dipoles align parallel to the spontaneous
polarization $P_s$ by diffusion of the mobile oxygen vacancy in cage motion.
Because of the relatively slow oxygen vacancy motion \cite{waser91bulk}, the
polarization directions of the aligned defect dipoles stay constant when the
direction of $P_s$ changes for short times. In this case, the defect dipoles
generate an internal electrical field ${\bf E}^i$ which stabilizes the domain
pattern by increasing the force constant for the reversible displacement of the
domain walls \cite{arlt93aging}. This relaxation model has been well developed
\cite{lohkaemper90internal}. It bears two insufficiencies, though, the time
dependence of aging is not reproduced and the absolute values of clamping
pressures are low \cite{lupascu06aging}. A second point of view about the role
of oxygen vacancies in aging is the formation of ionic space charges
\cite{dawber05physics} which was originally proposed to explain space charge
effects in BaTiO$_3$ single crystals \cite{wiliams65surface}. Ionic space
charges are well known for highly doped positive temperature coefficient
resistors based on BaTiO$_3$ \cite{ravikumar97space}. For aging the mobile
charge carriers move to charged domain faces or grain boundaries and compensate
polarization. This leads to an asymmetric charge distribution whereby a voltage
offset arises yielding the known shift or deformation of the ferroelectric
hysteresis \cite{pike95}. The clamping pressure on domain walls generated by
these space charges has not been treated mathematically for periodic domain
structures.

This paper describes quantitatively the formation of space charges in single domains of a
periodic structure and shows the development of the defect distribution inside
the domain. An estimate of bending and clamping pressures on domain walls and a
comparison to the orientational picture \cite{robels93domain} are given.
Electrostatic clamping of domain walls through the formation of space charges
is calculated to be two orders of magnitude stronger than clamping through
aligned defect dipoles for the same concentration of charge carriers.

The model is independent of the type of point defect, as long as a diffusion
constant can be assigned and the defect is charged. It can thus be equally well
applied to hopping of electronic carriers. The oxygen vacancy was chosen for
the numerical examples in order to be comparable to previous work, but does not
preclude a statement on the physical nature of the mobile carrier.

\section{\label{sec:generalmodel}General Model}

\begin{figure}[htbp]
\begin{center}
    \includegraphics[scale=.5]{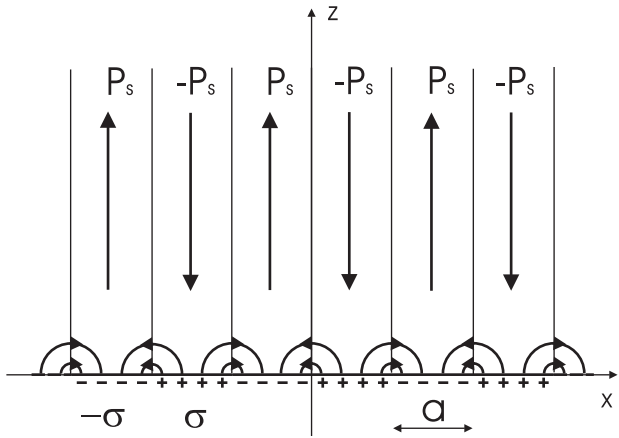}
    \caption{Scheme of a 2D-array of 180\°-domain walls crossing the grain boundary
     at a right angle. Straight arrows show the direction of the polarization and
     curved arrows the approximate directions of the local electric fields.}
    \label{domarray}
\end{center}
\end{figure}

In order to study the effect of migration of charge carriers on aging, we chose
a two-dimensional periodic array of domains cut by the grain surface, $z=0$,
perpendicular to the direction of spontaneous polarization which is along the
$z$ axis in Fig.~\ref{domarray}. This model configuration is well-known in the
physics of polarized media and was used for the study of equilibrium and
dynamic properties of ferromagnetic
\cite{Kittel1946,LandauElectrodynamicsContinuum} and ferroelectric
\cite{Mitsui1953,Fedosov1976} materials. We assume for simplicity an isotropic
material of the grain occupying the area $z>0$ characterized by the relative
dielectric constant $\varepsilon_f$. The dielectric material outside the grain
is assumed isotropic too and is characterized by the relative permittivity
$\varepsilon_d$. As we previously showed by finite element simulation, the
electric fields arising due to spontaneous polarization in a periodic
multi-domain grain of finite dimensions generate a nearly perfect periodic
pattern except for the very edges of the grain \cite{lupascu06aging}. We thus
consider the periodic domain array of Fig.~\ref{domarray} occupying the
semi-space $-\infty<x<\infty,\,z>0$ as a representative model for a
multi-domain grain of domain width $a$ much smaller than the typical grain
size.
Due to polarization, the domain faces at $z=0$ are alternatively charged with
the bound surface charge density $\sigma = |\mathbf{P}_s|, $ the spontaneous
polarization value. If the length of the domains $L$ along the $z$-axis is much
larger than their width $a$ along the $x$-axis, which is typically the case in
experiment, field lines are effectively closed at the same side of the grain
(see Fig.~\ref{domarray}). As a consequence, both components of the electric
field ${\bf E}^0(x,z)$ induced by the alternating surface charge exponentially
decrease towards the interior of the grain along the $z$-axis
\cite{lupascu06aging}. Migration of the charged defects driven by the field
${\bf E}^0(x,z)$ is then expected to occur in a volume near the grain surface.
The domains may therefore be considered infinitely long along the $z$-axis
without introducing a substantial error.

Let us now derive the equations of evolution for the charge and field
distributions in the considered system. At any time $t$, the electric field
${\bf E}(x,z,t)$ is determined by the charged faces of the domains and the
imbalanced charge density of free carriers $\rho(x,z,t)=q_f
\left[c(x,z,t)-c_0\right]$ through Gauss' law
\begin{equation}
\label{Gauss}%
\nabla{\bf E}=\frac{q_f}{
\varepsilon_0\varepsilon_f}(c-c_0)
\end{equation}
\noindent where  $c(x,z,t)$ is the local concentration of mobile carriers of
charge $q_f$, $c_0$ is the background concentration of the immobile charge
carriers of opposite polarity warranting total electroneutrality, and $\varepsilon_0$
is the permittivity of vacuum. In the initial state, the system is locally
neutral assuming $c(x,z,0)\equiv c_0$, the electric field ${\bf
E}(x,z,0)\equiv{\bf E}^0(x,z)$ is yet to be found.

In presence of an electric field and a gradient of concentration, the flow of
charge carriers is given by the sum of drift and diffusion contributions to the
particle current density:
\begin{equation}
\label{particle_current}%
{\bf s}=\mu c {\bf E} - D\nabla c
\end{equation}
\noindent where $\mu$ and $D$ are the mobility and diffusivity of charge
carriers, respectively. We assume, for simplicity, that the latter two
quantities are isotropic and connected by the Einstein relation, $D=\mu k
\Theta/q_f$ with $k$ the Boltzmann constant and $\Theta$ the absolute
temperature. Migration of charge carriers is governed by the continuity
equation:
\begin{equation}
\label{continuity}%
\partial_t c=-\nabla(\mu c {\bf E}) + D\triangle c  .
\end{equation}
For boundary conditions to the system of equations (\ref{Gauss}) and
(\ref{continuity}) we assume chemical and electrical isolation of the grain.
The first requirement means vanishing particle current
\begin{equation}
\label{boundary-chem}%
s_z=\mu c E_z - D\partial_z c=0,
\end{equation}
\noindent at the grain boundary, $z=0$. The second requirement
means vanishing total electric current,
\begin{equation}
\label{boundary-el}%
q_f(\mu c E_z - D\partial_z c)+ \varepsilon_0\varepsilon_f
\partial_t E_z=0,
\end{equation}
\noindent at $z=0$ which results in a constant value of the $z$-component of
the electric field at the grain boundary, $\partial_t E_z(x,0,t)=0$.

Eqs.~(\ref{boundary-chem},\ref{boundary-el}) together with
Eqs.~(\ref{Gauss},\ref{continuity}) complete the statement of the problem of
charge segregation in a ferroelectric grain. In the next section we will
observe how the system relaxes according to the equations of evolution
(\ref{Gauss},\ref{continuity}).

\section{\label{sec:solution}Solution of the equations of evolution}

In this section we first calculate the field ${\bf E}^0(x,z)$ in the virgin
state of the system before the process of charge segregation starts. Then we
formally solve equation (\ref{Gauss}) and find the total electric field ${\bf E}(x,z,t)$
for an arbitrary right-hand side.
Finally, using the latter result, we numerically solve equation
(\ref{continuity}), self-consistently describing drift and diffusion of the
mobile charge defects in the domain arrangement of Fig.~\ref{domarray}.

\subsection{\label{subsec:virgin}Electric field in the virgin state of a multi-domain grain}

\begin{figure}[htbp]
\begin{center}
    \includegraphics[scale=.5]{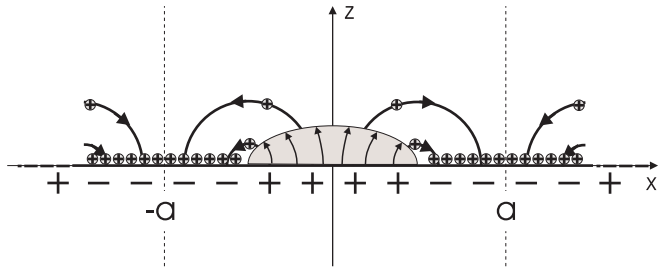}
    \caption{Scheme of expected charge redistribution induced by the local electric field within the
    central region $-a<x<a,\,z>0$, presenting repeating element of the periodic domain arrangement
    of Fig.~\ref{domarray}. Thin layers of positive charge carriers piled up at the negatively charged
    domain faces as well as a wide (shaded) area depleted of mobile charge carriers near the
    positively charged domain face are shown.}
    \label{centraldomain}
\end{center}
\end{figure}

To use the bilateral symmetry of the problem, the origin is chosen in the
center of the positively charged domain face as shown in
Fig.~\ref{centraldomain}.

The bound charge density of the domain faces is represented by an alternating
function \cite{LandauElectrodynamicsContinuum}
\begin{equation}
\label{face-charge}
\rho_b(x,z)=\sigma\delta(z)\sum_{n=-\infty}^{\infty}(-1)^n
\theta\left(\frac{a}{2}-an+x\right)
\theta\left(\frac{a}{2}+an-x\right),
\end{equation}
\noindent where $\delta (z)$ and $\theta  (x)$ are the Dirac $\delta $-function and the Heaviside unit
step function, respectively. The electrostatic potential induced by this bound charge is given by
the expression:
\begin{equation}
\label{face-potential-} \varphi_b(x,z)=-\frac{1}{2\pi
\varepsilon_0(\varepsilon_f+\varepsilon_d)}
\int_{-\infty}^{\infty}dx_0
\int_{-\infty}^{\infty}dz_0\,
\rho_b(x_0,z_0)
\ln\left[\left(\frac{x-x_0}{a}\right)^2 +
\left(\frac{z-z_0}{a}\right)^2 \right]
\end{equation}
\noindent in both areas $z\geq 0$ and $z<0$. The formula (\ref{face-potential-}) is simply a superposition
of the potentials generated by straight parallel charged lines located at the grain boundary $z=0$ between
the media with dielectric constants $\varepsilon_d$ and $\varepsilon_f$ \cite{LandauElectrodynamicsContinuum}.

The $z$-component of the electric field created by the bound charge, ${\bf E}^0=-\nabla \varphi_b$,
 may be directly calculated by substitution of
$\rho_b$, Eq.~(\ref{face-charge}), into Eq.~(\ref{face-potential-}) and
subsequent summation \cite{prudnikov86integrals} which results in the form
\begin{equation}
\label{E0z}
E^0_z(x,z)=\frac{2\sigma}{\pi\varepsilon_0
(\varepsilon_f + \varepsilon_d )}
\arctan{\left[\frac{\cos(\pi x/a)}{\sinh(\pi z/a)}\right]}
\end{equation}
\noindent valid for both media.

Direct calculation of the other field component,
$E^0_x=-\partial_x\varphi_b$, is more complicated because of slow convergence of
the appropriate series. Instead, $E^0_x$ may be calculated for $z>0$ from Gauss'
law $\nabla{\bf E}^0=0$, taking into account that, from
the symmetry of the problem, $E^0_x(0,z)=E^0_x(\pm a,z)=0$. Proceeding with
integration of the latter Gauss' equation over distance along the $x$-axis and
using the mentioned boundary conditions one finds the form
\begin{eqnarray}
\label{E0x}
E^0_x(x,z)=\frac{\sigma}{\pi\varepsilon_0 (\varepsilon_f + \varepsilon_d )}
\ln{\left[\frac{\cosh(\pi z/a)+\sin(\pi x/a)}
{\cosh(\pi z/a)-\sin(\pi x/a)}\right]}
\end{eqnarray}
\noindent valid inside and outside the grain.
Both field components exhibit periodic dependence along the $x$-axis, as
expected from the periodic domain arrangement, and exponentially decrease
at large distance from the charged surface $|z|\gg a$, as expected from
the previous finite element simulations \cite{lupascu06aging}.
We note here that the short range fields, Eq. (\ref{E0z},\ref{E0x}),
may be very large. For example, in BaTiO$_3$ the depolarization field amplitude
may be as high as $10^8 \rm \: V/m$ \cite{lupascu06aging}. The presence of high local
fields were confirmed at least partly in observations on the acceptor doped BaTiO$_3$,
where internal fields were experimentally found in the
range of $10^5 \rm \: V/m$ for up to 1 mol\% Ni-doping and temperatures below
80\°C \cite{arlt88internal}.

\subsection{\label{subsec:field}Electric field due to redistribution of charged defects}

At any arbitrary moment, the total electric field in the system may be conveniently
decomposed as ${\bf E}={\bf E}^0+{\bf E}^i$, where the field ${\bf E}^0$ is determined
by the charged faces of the domains, Eqs. (\ref{E0z},\ref{E0x}), and the field ${\bf E}^i$
is generated by the distribution of the charge density in the area $z>0$.
Thanks to periodicity and the bilateral symmetry of the initial
conditions, both the charge density and the electrostatic potential remain
periodic and bilaterally symmetric in the course of the charge redistribution,
as illustrated in Fig.~\ref{centraldomain}. This allows us to consider the region
$-a<x<a$ as a repetitive basic unit of the system and confine ourselves to
the consideration of processes in this area. To get a full description of the
electric field under these circumstances, it is sufficient to construct the
Green's function of the symmetrical Neumann problem in the mentioned region,
$G_s(x,z|x_0,z_0)$, so that the electrostatic potential induced by redistribution
of charged defects with $z_0>0$ may then be presented in a form \cite{jackson75classical}
\begin{equation}
\label{Pot-Green}
\varphi_i(x,z,t)=\int_{0}^{a}dx_0 \int_0^{\infty}dz_0\,
\rho(x_0,z_0,t) G_s(x,z|x_0,z_0),
\end{equation}
followed by the field expression ${\bf E}^i=-\nabla \varphi_i$.

The Green's function satisfies the equation
\begin{equation}
\label{Eq-Green}
\triangle G_s(x,z|x_0,z_0) = -\frac{1}{
\varepsilon_0\varepsilon_f}\delta (z-z_0) %\nonumber\\
%\times
\left[\delta(x-x_0) + \delta (x+x_0)\right]
\end{equation}
\noindent with boundary conditions $\partial_x G_s(x=\pm a ,z|x_0,z_0)=0$. The
latter requirement is a consequence of the constraint $E_{x}(\pm a,z)=0$
inherent to the chosen domain arrangement. Boundary conditions for the
electrostatic potential, Eq. (\ref{Pot-Green}),  on the interface between the
two media at $z=0$ \cite{jackson75classical} impose two additional boundary
conditions on the Green's function
\begin{eqnarray}
\label{Green-Boundaries}
G_s(x,-0|x_0,z_0)=G_s(x,+0|x_0,z_0)\nonumber\\
\varepsilon_d \partial_z G_s(x,-0|x_0,z_0)=\varepsilon_f \partial_z G_s(x,+0|x_0,z_0)
\end{eqnarray}

Using the fundamental solution of the 2D Poisson equation
\cite{jackson75classical} and taking into account periodicity of the problem
the solution of Eq.~(\ref{Eq-Green}) may be reduced to summation of a
series:

\begin{equation}
\label{Green-Series-}
G_s(x,z|x_0,z_0)=-\frac{1}{2\pi \varepsilon_0(\varepsilon_f + \varepsilon_d)}
\sum_{n=-\infty}^{\infty} \left\{\ln\left[
\left(\frac{x-x_0}{a}-2n\right)^2+
\left(\frac{z-z_0}{a}\right)^2\right]\right\}
+(x_0 \rightarrow -x_0)
\end{equation}
\noindent for the area $z<0$ and
\begin{eqnarray}
\label{Green-Series+}
G_s(x,z|x_0,z_0)=-\frac{1}{4\pi \varepsilon_0\varepsilon_f}
\sum_{n=-\infty}^{\infty} \left\{\ln\left[
\left(\frac{x-x_0}{a}-2n\right)^2+
\left(\frac{z-z_0}{a}\right)^2\right]\nonumber\right.\\
\left.
+\frac{\varepsilon_f - \varepsilon_d}{\varepsilon_f + \varepsilon_d}\ln\left[
\left(\frac{x-x_0}{a}-2n\right)^2+\left(\frac{z+z_0}{a}\right)^2\right]\right\}
+ (x_0 \rightarrow -x_0)
\end{eqnarray}
\noindent for the area $z>0$.

Because of slow convergence of this series it is more
convenient to perform summation for the derivatives $\partial_x
G_s$ and $\partial_z G_s$ and then to restore the function $G_s$
itself by integration using boundary conditions. This leads after
all to the solution of Eq.(\ref{Eq-Green})
\begin{equation}
\label{SymGreen-}
G_s(x,z|x_0,z_0)=-\frac{1}{2\pi\varepsilon_0(\varepsilon_f + \varepsilon_d)}
\ln\left[ \cosh\frac{\pi(z-z_0)}{a} -\cos\frac{\pi(x-x_0)}{a} \right]
+ (x_0 \rightarrow -x_0)
\end{equation}
\noindent for the area $z<0$ and
\begin{eqnarray}
\label{SymGreen+}
G_s(x,z|x_0,z_0)=-\frac{1}{4\pi\varepsilon_0 \varepsilon_f}
\left\{\ln\left[ \cosh\frac{\pi(z-z_0)}{a} -\cos\frac{\pi(x-x_0)}{a} \right]\right.\nonumber\\%
\left.+\frac{\varepsilon_f - \varepsilon_d}{\varepsilon_f + \varepsilon_d}
\ln\left[ \cosh\frac{\pi(z+z_0)}{a} -\cos\frac{\pi(x-x_0)}{a} \right]\right\}
+ (x_0 \rightarrow -x_0)
\end{eqnarray}
\noindent for the area $z>0$, which are periodic, bilaterally symmetric and
satisfy the proper boundary conditions. Now, from the expressions
(\ref{Pot-Green},\ref{SymGreen-},\ref{SymGreen+}), the components of the
electric field induced by the redistribution of charged defects may be obtained.
It is easy to verify that the total electric
field satisfies the boundary condition $E_x(x=\pm
a,z)=0$ for any bilaterally symmetric charge density $\rho(x,z,t)$.

\subsection{Numerical solution of the evolution equations}

Having solved equation (\ref{Gauss}) explicitly allows for the implementation
of a simple direct Euler scheme for numerical treatment of the problem. Space
and time will be discretized. At every time step, the change in the carrier
concentration will be calculated from the previous values of the concentration
and the electric fields using Eq.~(\ref{continuity}). Then, the updated values
of the field will be calculated directly from the updated values of the
concentration using Eq.~(\ref{Pot-Green}). The calculation is repeated until
convergence. Taking into account the bilateral symmetry of the problem, it is
sufficient to consider the charge redistribution within the area $0<x<a$.

We first introduce dimensionless variables which is helpful for the following
numerical analysis and reveals those parameters of the system which are
relevant to the relaxation process. Dimensionless coordinates are naturally
introduced as $X=x/a$ and $Z=z/a$. The dimensionless field ${\bf F}={\bf
E}/E^{\ast}$ is expressed in units of the characteristic value
$E^{\ast}=\sigma/2\varepsilon_0\varepsilon_f$. The system reveals two
characteristic time scales: the drift time $\tau_{\mu}=a/\mu E^{\ast}$ and the
diffusion time $\tau_D=a^2/D$. For the typical parameters involved, $\tau_D \gg
\tau_{\mu}$ therefore we will introduce dimensionless time as $T=t/\tau_{\mu}$.
The concentration of defects is now reduced to $n(X,Z,T)=c(x,z,t)/c^{\ast}$
with the characteristic value $c^{\ast}=\sigma/2aq_f$. The latter has the
physical meaning of a concentration of defects on an area $a^2$, which
completely neutralizes the bound charge $\sigma$ at the domain faces. The
reduced initial concentration $n_0=c_0/c^{\ast}$ measures whether the density
of defects is high or low with respect to the charge compensation
concentration.

The continuity equation (\ref{continuity}) now acquires the form
\begin{equation}
\label{continuity2}%
\partial_t n=-n(n-n_0)- {\bf F}\nabla n + \beta\triangle n
\end{equation}
\noindent where all differentiations are performed with respect to the
dimensionless variables. The parameter $\beta = \tau_{\mu}/\tau_D \ll 1$
characterizes a weak contribution of diffusion to the migration of defects in
ferroelectrics. It is now seen from Eq.~(\ref{continuity2}) that only two
composed parameters, $n_0$ and $\beta$, control the relaxation process.

Though the parameter $\beta$ may be rather small, it cannot be neglected as is
clearly seen from the boundary condition for the particle current,
Eq.~(\ref{boundary-chem}), taken in a dimensionless form
\begin{equation}
\label{boundary-chem-dim}%
n F_y - \beta\partial_y n=0,\,\,\,\,Z=0,
\end{equation}
\noindent otherwise this boundary condition is not compatible with the initial
conditions. The finite value of $\beta$ means compensation of the drift
contribution to the current by the diffusion contribution at the grain boundary
and this way defines the structure of a thin layer of charged defects piling up
at this boundary.

Eq.~(\ref{continuity2}) is supplemented by expressions for the dimensionless
field ${\bf F}={\bf F}^0+{\bf F}^i$ which can be easily derived from Eqs.
(\ref{E0z},\ref{E0x}) and (\ref{Pot-Green},\ref{SymGreen-},\ref{SymGreen+}), namely,
\begin{eqnarray}
\label{DimField0}
F^0_{x}(X,Z)=\frac{1}{\pi} \frac{2\varepsilon_f}{\varepsilon_f+\varepsilon_d} \ln{\left[\frac{\cosh\pi Z+\sin\pi
X}{\cosh\pi Z -\sin\pi X}\right]}\nonumber\\
F^0_{z}(X,Z)=\frac{2}{\pi}\frac{2\varepsilon_f}{\varepsilon_f+\varepsilon_d}\arctan{\left[\frac{\cos\pi X}{\sinh\pi Z}\right]}
\end{eqnarray}
\noindent and
\begin{equation}
\label{DimField} F^i_{x,z}(X,Z,T)=\int_{0}^{1}dX_0
\int_0^{\infty}dZ_0\,f_{x,z}(X,Z|X_0,Z_0)
\left[n(X_0,Z_0,T)-n_0\right]
\end{equation}
\noindent where the kernels in this integral are presented by the functions
\begin{eqnarray}
\label{kernels-}
f_{x}(X,Z|X_0,Z_0)=\frac{\varepsilon_f }{2(\varepsilon_f+\varepsilon_d)}
\frac{\sin\pi (X-X_0)}{\cosh\pi (Z-Z_0)-\cos\pi (X-X_0)}
+ (X_0 \rightarrow -X_0),\nonumber\\
f_{z}(X,Z|X_0,Z_0)=\frac{\varepsilon_f }{2(\varepsilon_f+\varepsilon_d)}
\frac{\sinh\pi(Z-Z_0)}{\cosh\pi (Z-Z_0)-\cos\pi (X-X_0)}
+ (X_0 \rightarrow -X_0)
\end{eqnarray}
\noindent for $Z<0$ and by functions
\begin{eqnarray}
\label{kernels+}
f_{x}(X,Z|X_0,Z_0)=
\frac{1}{4}\left[\frac{\varepsilon_f-\varepsilon_d}{\varepsilon_f+\varepsilon_d}
\frac{\sin\pi(X-X_0)}{\cosh\pi (Z+Z_0)-\cos\pi (X-X_0)} \right.\nonumber\\
\left. +\frac{\sin\pi(X-X_0)}{\cosh\pi (Z-Z_0)-\cos\pi (X-X_0)}
+ (X_0 \rightarrow -X_0)\right],\nonumber\\
f_{z}(X,Z|X_0,Z_0)=
\frac{1}{4}\left[\frac{\varepsilon_f-\varepsilon_d}{\varepsilon_f+\varepsilon_d}
\frac{\sinh\pi(Z+Z_0)}{\cosh\pi (Z+Z_0)-\cos\pi (X-X_0)} \right.\nonumber\\
\left. +\frac{\sinh\pi(Z-Z_0)}{\cosh\pi (Z-Z_0)-\cos\pi (X-X_0)}
+ (X_0 \rightarrow -X_0)\right]
\end{eqnarray}
\noindent for $Z>0$.

Since the system remains electrically neutral within the domain of integration
during the redistribution of defects, arbitrary constants may be added to the
kernels (\ref{kernels-},\ref{kernels+}) without changing the results of
integration in Eqs.~(\ref{DimField}). This property is used in the
numerical procedure to facilitate the conversion of the integrals in
Eqs.~(\ref{DimField}).

As an example, we now consider the aging process in $\rm BaTiO_3$. For the
numerical simulations, the material parameters of $\rm BaTiO_3$ at room
temperature are taken from Wernicke and Jaffe et al.~\cite{wern75,jaff71},
namely, $P_s=2.71\cdot 10^{-5} \rm \: C/cm^2$, $\varepsilon_f = 170$,
$\mu=1.73\cdot 10^{-20} \rm \: m^2/Vs$, $a=0.5\rm \: \mu m$ and $q_f$ twice the
elementary charge, implying positively charged oxygen vacancies as mobile
defects. For the dielectric medium between ferroelectric grains we take the
same but non-polarized material with $\varepsilon_d = 170$. This yields
$c^{\ast} = 1.69\cdot10^{18} \rm \: cm^{-3}$, $\tau_{\mu} = 1.61\cdot10^{5} \rm
\: s$, $\tau_{D} = 1.14\cdot10^{9} \rm \: s$. As was shown in one dimensional
simulations \cite{lupascu06aging}, the parameter $\beta <<1$ has no effect on
the dynamics of the relaxation. The only physical characteristic depending on
$\beta <<1$ is the thickness of the positively charged layer of defects piling
up at the negative face of the domain. To make this layer visible in figures
and to avoid numerical problems invoked by the strong gradients of the defect
density we take the value $\beta=5\cdot 10^{-2}$ instead of the actual ratio
$\tau_{\mu}/\tau_D=1.4\cdot 10^{-4}$ for our simulations.
\begin{figure}[htbp]
\begin{center}
\includegraphics[scale=.90]{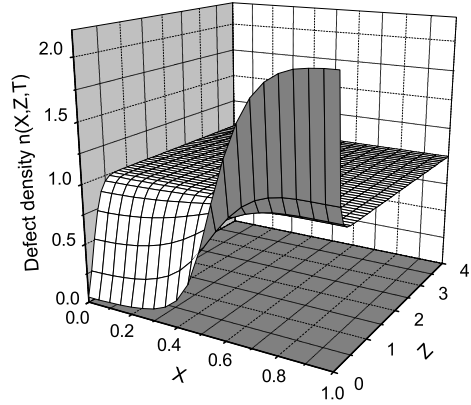}
    \caption{Distribution of oxygen vacancies $c_{V_o}(X,Z)$ over the
    reference area $0<X<1, 0<Z<4$ at time $T=0.05$ for an initial
    concentration of defects $c_0=c^{\ast} = 1.69\cdot10^{18} \rm \: cm^{-3}$.}
    \label{relief}
\end{center}
\end{figure}

A snapshot of the development of the defect concentration profile over the
reference area $0<X<1,\, 0<Z<4$ starting with the background defect
concentration $n_0 =1$ is presented in Fig.~\ref{relief} for the moment $T=0.05$.
A wide depleted zone forms near the positively charged face
at $0<X<0.5, Z=0$ and a very thin excess charge layer of high concentration
near the negatively charged face at $0.5<X<1, Z=0$.
\begin{figure}[htbp]
\begin{center}
    \includegraphics[scale=.67]{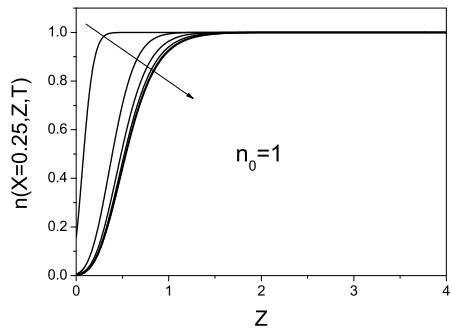}
    \caption{Defect concentration profile along the line $X=0.25$
     for a succession of times $T=0.1,\,1,\,2,\,3,\,4,\,5$ (from left
    to right) for the initial concentration of defects $c_0=n_0 \cdot c^{\ast}$. }
    \label{cross-left}
\end{center}
\end{figure}

\begin{figure}[htbp]
\begin{center}
    \includegraphics[scale=.62]{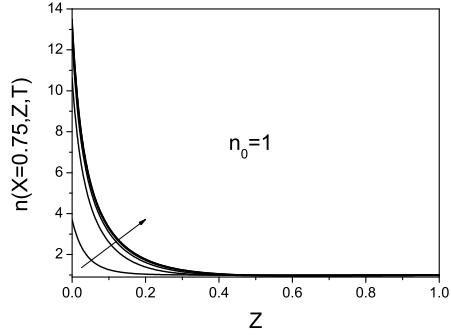}
    \caption{Defect concentration profile along the line $X=0.75$
     for a succession of times $T=0.1,\,1,\,2,\,3,\,4,\,5$ (from left
    to right) for an initial concentration of defects $c_0=n_0 \cdot c^{\ast}$.}
    \label{cross-right}
\end{center}
\end{figure}

\begin{figure}[htbp]
\begin{center}
    \includegraphics[scale=.6]{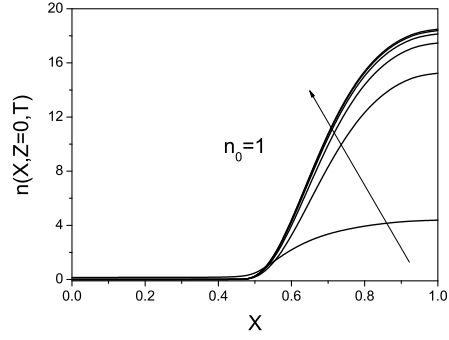}
    \caption{Defect concentration profile along the line $Z=0$
     for a succession of times $T=0.1,\,1,\,2,\,3,\,4,\,5$ (upwards)
     for the initial concentration of defects $c_0=n_0 \cdot c^{\ast}$.}
    \label{cross-front}
\end{center}
\end{figure}

The structural difference of these two space charge areas is better seen in
Figs.~\ref{cross-left} and \ref{cross-right} presenting vertical cross sections
of the concentration profile along the lines $X=0.25$ and $X=0.75$,
respectively. A succession of snapshots of the concentrations along the line
$X=0.25$ (Fig.~\ref{cross-left}) exhibits the evolution of the charge defect
density near the positive face of the domain. The profile positions at the
moments $T=4$ and $5$ cannot be discerned any more which indicates saturation
at time $T\simeq 5$ (corresponding to $t\simeq 8\cdot 10^5 \rm \: s$). The
characteristic width of this zone in the final relaxed state is of the order of
unity. The defects piling up near the negative face of the domain form a much
thinner layer of a characteristic width of the order of $\beta$ as is
represented by concentration profiles along the line $X=0.75$ in
Fig.~\ref{cross-right}. The final relaxed state is reached also at about
$T\simeq 5$. The corresponding evolution of the front cross section of the
concentration profile along the line $Z=0$ shown in Fig.~\ref{cross-front}
exhibits saturation at about $T\simeq 5$, too.

\begin{figure}[htbp]
\begin{center}
    \includegraphics[scale=.62]{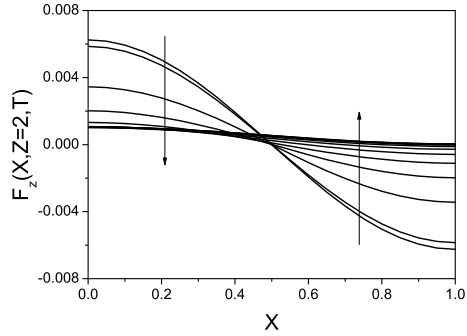}
    \caption{The electric field component $F_z$ plotted along the line
$Z=2$ for a succession of times $T=0,\,0.1,\,1,\,2,\,3,\,4,\,5,$
$\,6,\,7,\,8,\,9,\,10$ in a system with the initial concentration of defects
$c_0=c^{\ast}$. The arrows show the direction of evolution.}
    \label{cross-fieldEy}
\end{center}
\end{figure}

In our model, drift-dominated migration of the charged defects is caused by
local electric fields near the charged faces of a grain. This migration process
only stops, if either no mobile defects remain in the area where fields are
present or there is no remaining field in the area where the defect
concentration is not zero. The process of field compensation due to defect
migration is exemplified by the evolution of the electric field component
$F_z=F^0_z+F^i_z$ at the line $Z=2$ represented in Fig.~\ref{cross-fieldEy}.
$F^i_z$ saturates at the values opposite to the local values of the initial
electric field $F^0_z$ determined by the bound surface charge.
Relaxation leads to an energy minimum where the system will resist
any change of the domain wall positions. The final distribution of free charges
then determines the equilibrium domain configuration of the system. For an
effectively low mobility of the free charge carriers the transition above the
Curie point will not readily rearrange the charge carrier configuration due to
thermal excitation. The defect charge density then determines the subsequent
domain configuration after re-cooling the sample to low temperature.
Experimentally it is observed that the original domain configuration is
reproduced to a great extent \cite{zhang05insitu,zhang06aging,hennings}.

\section{\label{sec:pinningforce}Forces exerted upon a domain wall}

From the known development of the charge density and the electric field in our
model, the time dependent forces exerted upon domain walls can be evaluated.
Using the general formula derived by Nechaev et al. \cite{nechaev90effect} and
taking into account only electrostatic contributions to the energy one can
obtain the local pressure $f$ exerted upon a wall. For a straight rigid wall
considered here, one finds $f=2{\bf P E}$ where $\bf P$ and $\bf E$ are the
local values of spontaneous polarization and electric field, respectively. This
relation is reduced, in the geometry of Fig.~\ref{domarray}, to $f=2P_s E_z$
(note that, in the case of the same arrangement of the 90\°-domain walls, the 
$\sigma$ would merely decrease by a factor of $\sqrt{2}$ and the force by a factor
of 2, the configuration and results are otherwise identical).

Although only one end of the domains is present in the mentioned geometry of
Fig.~\ref{domarray} it is obvious that similar segregation of the charged
defects occurs on the other end of the domain, too. This results in the
antisymmetric force of opposite sign exerted upon the domain wall on the other
end of the domain yielding a total force equal to zero. This force cannot move
the domain wall as a whole or prevent its motion but it may lead to bending of
the wall violating our assumption of rigid straight domains. This is frequently
encountered in real systems. Domains forming needle tips near external
interfaces are commonly observed \cite{salje96mesoscopic,shur00formation}. 
In this case, part
of the compensation arises within the bulk and not only right at the grain
interface. The final defect distributions will be different from the case
calculated here, but the essential effect of bending will remain the same. Our
model of drift of free charge carriers also supports a coalescence of domains
rather than their splitting. Without any further details included in the model,
it contradicts the experimental observations of Ikegami ad Ueda of domain
splitting during aging \cite{ikegami67mechanism}.

\begin{figure}[htbp]
\begin{center}
    \includegraphics[scale=.62]{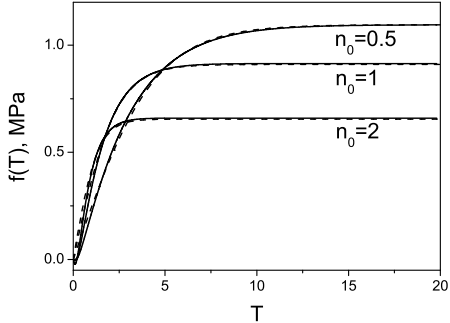}
    \caption{Bending pressure $f$ as a function of time $T=t/\tau_{\mu }$
at room temperature for $\rm BaTiO_3$ is plotted for three different
sample concentrations of oxygen vacancies
$c_0 = n_0 \cdot c^{\ast}$ with $n_0 = 0.5,\,1$ and $2$
(solid lines). Dashed lines show fitting of the pressure by the function
$f_0 \tanh{(\alpha n_0 T/2)}$ with parameters
$f_0=1.095,\,0.91,\,0.66 \rm \: MPa$ and $\alpha =0.91,\,0.86,\,0.76$
for $n_0 = 0.5,\,1$ and $2$, respectively.
}
    \label{clamping}
\end{center}
\end{figure}

The evolution of the bending pressure $f(T)$ averaged over half the domain wall
length, assumed as long as $L=20a$, is shown in Fig.~\ref{clamping} for three
different values of the initial background concentration of defects. It is seen
that systems with smaller concentrations need an inversely longer time to
relax. For the system with $n_0=1$ it takes about $T\simeq 5$ while for the
system with $n_0=0.5$ this time is roughly doubled. All curves can be well
fitted by the exponential form $f_0 \tanh{(\alpha n_0 T/2)}$ where the
parameters $f_0\simeq 1$~MPa and $\alpha\simeq 1$ and slowly increase when
$n_0$ decreases. A reliable simulation of defect concentrations smaller than
$n_0=0.5$ is impossible on the chosen template $(0<X<1,\,0<Z<4)$ since in this
case migration involves defects from a wider area in order to compensate the
bound charge at the domain faces.

\begin{figure}[htbp]
\begin{center}
    \includegraphics[scale=0.62]{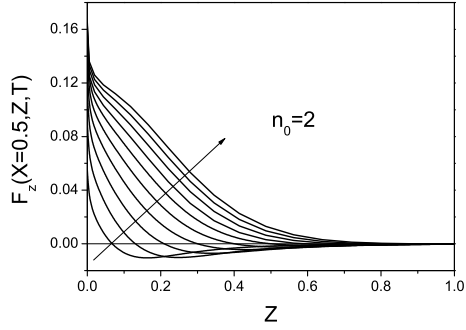}
    \caption{Snap-shots of the distribution profile of the field component $F_z$
along the domain wall for the succession of times
$T=0.1,\,0.2,\,0.3,\,0.4,\,0.5,\,0.6,\,0.7,\,0.8,\,0.9,\,1$ (upwards) for $\rm BaTiO_3$
at room temperature for an oxygen vacancy concentration $c_0=n_0 \cdot c^{\ast}$.}
    \label{pressure-t}
\end{center}
\end{figure}

One more general feature of time dependencies of the bending pressure in
Fig.~\ref{clamping} is worth discussion. All the curves exhibit a small region
at small times where the value of pressure is negative. This is not an artefact
of the numerical discretization procedure but has a physical meaning. Indeed at
any time, the characteristic width of the positive space charge zone near the
domain face $0.5<X<1, Z=0$ is of the order of $\beta $ which follows from the
boundary condition, Eq.~(\ref{boundary-chem}). On the other hand, at the very
beginning of charge defect migration, the characteristic width of the negative
space charge zone in the area $0<X<0.5, Z>0$ is less than $\beta$. This means
that a negative value of the field component $F_z$ prevails at the domain wall
at this stage. This is confirmed by the dependence of $F_z$ on position $Z$ for
different times as presented in Fig.~\ref{pressure-t}.

The above considered bending force does not directly describe the aging
phenomenon as long as rigid straight domain walls are retained. In fact, the total
force exerted upon the walls remains equal to zero during the defect redistribution
if both ends of the domains are taken into account. 
Nevertheless,
the loss of domain wall mobility characteristic of aging may be captured in
this model, too. Indeed, the segregation of charge carriers in the fixed domain
framework of Fig.~\ref{domarray} leads to the relaxation of the energy of the
electrostatic depolarization field. The decrease of this energy per unit length
of domain wall measures the clamping pressure preventing the displacement of
the wall from the energy minimum:
\begin{equation}
\label{clamping-form}%
P_{cl}(z,t)=\frac{\varepsilon_0 \varepsilon_f}{2a} \int_0^a dx
\left( {\bf E}^0(x,z)^2 -{\bf E}(x,z,t)^2 \right)
\end{equation}

\begin{figure}[htbp]
\begin{center}
    \includegraphics[scale=0.62]{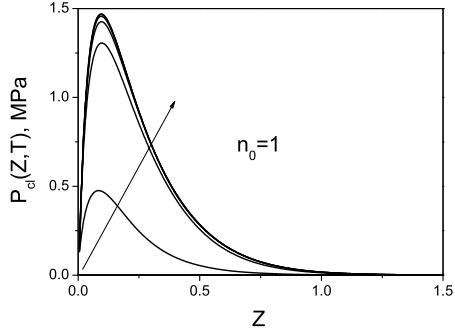}
    \caption{Snap-shots of the clamping pressure distribution
along the domain wall for the succession of times
$T=0,\,0.1,\,1,\,2,\,3,\,4,\,5$ (upwards) for $\rm BaTiO_3$ at room temperature for an
oxygen vacancy concentration $c_0 = n_0 \cdot c^{\ast}$. }
\label{clamping-figure}
\end{center}
\end{figure}

\begin{figure}[htbp]
\begin{center}
    \includegraphics[scale=0.62]{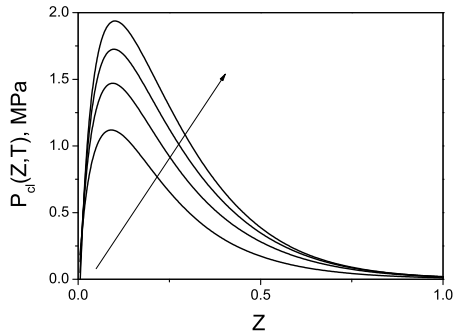}
    \caption{The saturated clamping pressure distribution
along the domain wall for $\rm BaTiO_3$ at room temperature for the
oxygen vacancy concentrations $c_0 = n_0\cdot c^{\ast}$ with
$n_0 = 0.5,\,1,\,1.5$ and  $n_0 = 2$ (upwards).}
\label{comparePCL}
\end{center}
\end{figure}

The dependence of this pressure along the length of the wall is shown in
Fig.~\ref{clamping-figure} for a succession of times. The magnitude of the
pressure saturates as expected at about a time $T\simeq 5$ for a defect
concentration of $c_0 = c^{\ast}$. The corresponding peak value of the pressure
around 1.5~MPa is comparable with the average bending pressure at the wall,
Fig.~\ref{clamping}.
The magnitude of the saturated pressure increases monotonously with the defect
concentration $c_0$ as is seen from Fig.~\ref{comparePCL}.

\begin{figure}[htbp]
\begin{center}
    \includegraphics[scale=0.5]{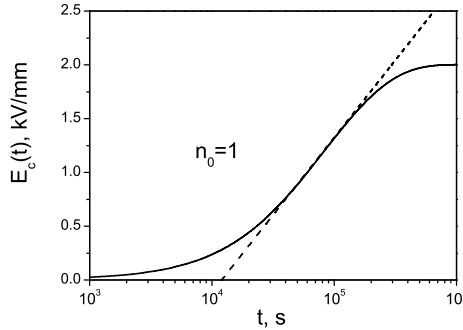}
    \caption{Coercive field due to charged defect migration
as a function of time  (solid line) for the oxygen vacancy
concentration $c_0 = n_0\cdot c^{\ast}$. Dashed line shows fitting with logarithmic
dependence for intermediate times.}
\label{CoerciveField}
\end{center}
\end{figure}

The irreversible migration of charged defects entails growing immobilization
of the domain walls and, consequently, enhancement of the coercive field, $E_c$. To
estimate this effect one should compare the pressure $\sim P_s \mathcal{E}$ exerted
by the external field, $\mathcal{E}$, upon a domain wall  with the clamping pressure,
Eq. (\ref{clamping-form}), averaged over the domain wall length, $L$. This results in
the following estimate for the coercive field
\begin{equation}
\label{coercive-field}%
E_c(t)=\frac{2}{P_s L} \int_0^{L/2} dz
P_{cl}(z,t)
\end{equation}
where integration over the half-length of the wall accounts for the other end
of the domain. Evaluation of the time-dependent coercive field assuming the
typical length of the domain wall $L=20a$ obtains a characteristic value
of $E_c\simeq 1 \rm \: kV/cm$ which is of the order of the coercive field in
unaged bulk samples of $\rm BaTiO_3$ \cite{zhang05insitu,zhang06aging}. 
In fact, the magnitude of the clamping pressure and, consequently, the value of
the coercive field may be substantially larger then it was estimated using 
Eq. (\ref{coercive-field}). Firstly, the peak value of the pressure has to be 
approximately doubled if one takes into account the reduction of the energy of 
electrostatic field in the dielectric material outside the grain which is approximately 
the same as in the ferroelectric area assuming $\varepsilon_d =\varepsilon_f$. Secondly, 
the consideration of the anisotropy of the dielectric constant is expected to scale up 
the pressure together with the energy gain by the factor of 
$\sqrt{\varepsilon_a/\varepsilon_c}$ which is about 6 for $\rm BaTiO_3$.
Finally, values of few $\rm\: MPa$ are expected for the average clamping pressure
at the domain wall and the values of few $\rm\:kV/mm$ are expected for the
coercive field due to charge carrier migration which is in agreement with the
characteristic values observed on the aged samples of $\rm BaTiO_3$
\cite{zhang05insitu,zhang06aging}. Accordingly, the coercive field, 
Eq. (\ref{coercive-field}), multiplied by the factor of $12$ is shown in 
Fig.~\ref{CoerciveField} in physical units to compare with known experimental data. 
The dashed line shows that the time behavior of $E_c$ mimics logarithmic time dependence 
for durations less then a few $\tau_{\mu }$.    

One more essential factor which brings about enhancement of the coercive field is that 
the minimum energy of the system will further substantially decrease if domain wall
bending is allowed contributing to the increase of the clamping pressure,
Eq. (\ref{clamping-form}). This mechanism is, however, beyond the consideration in our
model of rigid walls. 

The above obtained values are much larger than typical magnitudes of clamping pressure
arising due to dipole re-orientation \cite{robels93domain}. Indeed, for
uniformly aligned dipoles in the latter mechanism, the dipole moments exert
upon the domain wall a clamping pressure $P_{or}\simeq c_0 E_z d$ where the
dipole moment $d=q_f l/2$ with the dipole length of $l=4\cdot10^{-10} \rm \:
m$. For the material parameters assumed in the above estimations  and
$c_0=c^{\ast}$ this results in the peak value of the clamping pressure
$P_{or}=9.7 \rm \: kPa$ which is two orders of magnitude smaller than that in
the drift mechanism.

A common feature of these two aging mechanisms is that both dipole
re-orientation and defect migration occur in those areas where a depolarization
electric field is present. In respect thereof these mechanisms can be
classified neither as volume nor as boundary ones as was suggested in the
recent works by Zhang and Ren \cite{zhang05insitu,zhang06aging} but rather as
geometry dependent. Indeed, in the two-dimensional periodic array of domains
considered here the depolarization field is present only near the grain
boundaries causing both defect drift and dipole re-orientation only in this
area. On the other hand, in the single domain state of a Mn-doped $\rm BaTiO_3$
single crystal observed in Ref.~\cite{zhang06aging} a one-dimensional geometry
is virtually realized  where the depolarization field is present in the whole
sample \cite{lupascu06aging} and invokes both dipole re-orientation and defect
drift in the whole volume.

\section{\label{sec:conclusions}Conclusions}

In this work, we have considered migration of charged defects as a possible reason
for aging in ferroelectrics. The model is based on two main assumptions: 1) existence
of mobile carriers of ionic or electronic nature in the bulk material and 2) presence
of strong local depolarization fields due to bound charges at the domain faces. The
first assumption is based on direct measurements of the conductivity in perovskites
\cite{waser91bulk}, the second one was at least partly confirmed in observations of
Ref. \cite{arlt88internal}. Solving self-consistently the drift-diffusion equation
together with the Gauss equation for the fixed two-dimensional domain array
\cite{Kittel1946,LandauElectrodynamicsContinuum,Mitsui1953,Fedosov1976}
reveals gradual formation of space charge zones compensating the field generated by charged
domain faces. Charged domain walls, which are tip-to-tip or tail-to-tail configurations 
of the polarization in adjacent domains, are electrically equivalent to our model. The 
biggest difference arises due to the fact that charged domain walls are often observed as needle
tip domains in single crystals \cite{shur00formation}. The geometry is thus considerably 
different from the model of parallel domain walls presented in our paper, where
only periodic straight domain configurations are captured.

The process of charge defect migration is accompanied by the reduction of the energy of the
electrostatic depolarization field which leads to the energy minimum where the system
will resist any change of the domain pattern. The characteristic time of this relaxation
depends on the doping and is typically about
$5\cdot\tau_{\mu}\simeq 8\cdot 10^5 \rm \: s \simeq 9$ days, where $\tau_{\mu}$ is a time
of drift over the distance of domain width. That is why, after aging, a clamping force
at a domain wall arises if an external electric field attempts to shift the domain
wall from its initial position. This force may be estimated from the calculated
energy gain due to the reduction of the depolarization field. The peak value of the clamping
pressure is in the range of $1\div 10 \rm\: MPa$ but the pressure is distributed very
inhomogeneously along the domain wall concentrating near the domain ends. Nevertheless,
the total value of the clamping force at the domain wall results in the characteristic
coercive field of few $\rm\:kV/mm$ which is comparable with that observed on the aged
samples of Mn-doped $\rm BaTiO_3$ \cite{zhang06aging}.

Clamping pressures on domain walls in the presented two-dimensional model are considerably
lower than in the uniaxial case \cite{lupascu06aging} and approach macroscopically
observable values. They are two orders of magnitude larger than in the picture
of defect dipole re-orientation \cite{robels93domain} and are thus a plausible
mechanism for aging in ferroelectrics.
In contrast to the one-dimensional case with only one
characteristic value of electric field, $E_d=P_s/\varepsilon_f \varepsilon_0$,
treated earlier \cite{lupascu06aging} the two-dimensional model exhibits seemingly a
wide spectrum of characteristic times according to the position-dependent
values of the electric field ${\bf E}(x,y)$. This allows one to expect a time
dependence of the clamping pressure in a two-dimensional array of domains
different from the one-dimensional case \cite{lupascu06aging}. Nevertheless,
comparing time evolution of the field and defect concentration in Ref.
\cite{lupascu06aging} with Figs.
(\ref{cross-left},\ref{cross-right},\ref{cross-front},\ref{cross-fieldEy}) one
observes a striking similarity between them. We are thus concerned with a
single characteristic time constant $\tau_r = \tau_{\mu}/n_0$ characterizing
the relaxation of the system. This time is independent of the width of the
domains, $a$. In fact, $\tau_r =\varepsilon_f \varepsilon_0 /\lambda$ with
$\lambda=q_f c_0 \mu$ being the conductivity of the material is the Maxwell
relaxation time which only depends on the mobility and local concentration of
the mobile carriers. This in turn means that a distribution of grain sizes in
the material and accordingly a distribution of domain sizes does not entail a
distribution of characteristic relaxation times. The logarithmic time
dependence of the dielectric constant during aging yet remains to be explained.

A crucial parameter for the plausibility of the time scale in our simulations
is the mobility of charged species in a ferroelectric material. The mobility of
oxygen vacancies considered is still a highly disputable issue. The activation
barrier for this ionic defect is usually estimated in the range of 0.9-1.1 eV
in both experimental works and first principle calculations
\cite{TagantsevReview,MeyerWaser2005,DamjanovicReview,Erhart2007} which makes
the migration of oxygen vacancies over the distance of the order of the domain
width $\simeq 0.5 \rm\: \mu m$ most unlikely. On the other hand, the
estimations of the mobility in the
Refs.~\cite{waser91bulk,dawber05physics,Dawber2000-1,Jiang2002} are similar to
or higher than that given in Refs.~\cite{wern75,jaff71} which we used for
simulations in our study. We would like to stress here therefore that the
nature of the charge carriers plays no important role for the model presented.
These may be also electronic carriers as was suggested in
Refs.~\cite{MeyerWaser2005,Warren1995}. In any case our input parameters agree
with direct measurements of the conductivity of perovskites indifferent to the
nature of the charge carriers~\cite{waser91bulk}.

It is evident that any real system will contain more than one mobile charge
carrier. In case their mobilities or concentrations are considerably different,
the final distribution of defects of the more mobile / more frequent carrier
will determine the field environment for the drift of the second carrier
as it was discussed in Ref.~\cite{MeyerWaser2005}. The solutions from the present
calculation would have to be taken as starting condition and iteratively the
final solution could be found. In case of equal
mobilities and concentrations, a coupled system of equations has to be solved
which is the issue of forthcoming work. Similarly the local potential wells for
the domain wall which determine the dielectric constant will be given in a
future publication.

\section{\label{sec:acknowledgement}Acknowledgement}

Discussions with Karsten Albe, Nina Balke, Dietmar Gross, Valeriy Ishchuk, Hans Kungl,
Ralf M\"{u}ller, Hermann Rauh,  and J\"{u}rgen R\"{o}del and the support by the
Sonderforschungsbereich 595 of the Deutsche Forschungsgemeinschaft are gratefully
acknowledged.

\bibliographystyle{plain}
\bibliography{apssamp}

\end{document}